# Bismuthylene Monolayer: A Novel Quantum Spin Hall Insulator with Large Band Gaps

Run-wu Zhang,[a] Chang-wen Zhang,[a],* Wei-xiao Ji,[a] Ping Li,[a] Pei-ji Wang[a]

[a] School of Physics and Technology, University of Jinan, Jinan, Shandong, 250022, People's Republic of China

**Abstract:**

By means of first-principles calculations, we predict a new 2D QSH insulator in the porous allotrope of Bismuth monolayer, bismuthylene, its dynamics stability being confirmed by phonon spectrum and molecular dynamics simulations. The analyses of electronic structures reveal that it is a native QSH state with a gap much as large as 0.29 eV at the Γ point, which is larger than the buckled (0.2 eV) and flattened (0.2 eV) bismuth, $Bi_4Br_4$ (0.18 eV), as well as stanene (0.1 eV), also more stable energetically than these systems. Interestingly, the bismuthylene has tunable band gaps and nontrivial band topology under strains within -6 – 5 % and electric fields up to 0.8 eV/Å. Furthermore, a tight-binding model is constructed to explain the low-energy physics behind band topology induced by spin-orbit coupling. We also propose a quantum well by sandwiching bismuthylene between two BN sheets and reveals that this structure remains topologically nontrivial with a sizeable band gap. This findings on QSH effect of bismuthylene provide a viable platform in new generation of dissipationless electronics and spintronics devices.

\_\_\_\_\_\_\_\_\_\_\_\_\_\_\_\_\_\_\_\_\_\_\_\_\_\_\_\_\_\_\_\_\_\_\_\_\_\_\_\_\_\_

* Correspondence author: ss_zhangchw@ujn.edu.cn

Two-dimensional (2D) topological insulators (TIs), also known as quantum spin Hall (QSH) insulators, have attracted lots of attentions in recent days, owing to their exotic physical properties. [1] The unique characteristic of 2D TI is generating a gapless edge state inside bulk gap, in which the edge state is topologically protected by time-reversal symmetry (TRS) and more robust against backscattering than the 3D TI, making 2D TIs better suited for coherent spin transport related applications [1-3]. The robustness of edge states against nonmagnetic impurities makes 2D TIs better suited for coherent spin-transport related applications. Motivated by the QSH model in graphene proposed by Kane and Mele [4,5], several group-IV thin films such as Si [6], Ge[7,8], Sn[9,10,11], and Pb[12], have been put forward theoretically to harbor QSH effect, but the experimental demonstration of the existence of topological spin transport channels is only limited to HgTe/CdTe[13,14] and InAs/GaSb[15,16] quantum wells at very low temperatures and ultrahigh vacuum due to weak spin-orbit coupling (SOC). To expand and advance the practical application of 2D films, it is essential to search for new 2D TIs of high working temperatures to overcome the thermal disturbance for future electronic device applications.

Bismuth (Bi), the counterpart of carbon among group-V honeycomb lattice, is known for its stronger SOC which can drive nontrivially topological states. Thus, the compounds containing element Bi are among the most promising candidates, examples including monolayers Bi (111),[17,18] III-Bi,[19] functionalized Bi,[20,21] halogenated GaBi,[22] and so on. Especially, Drozdov *et al.*[23] report on the 1D topological edge states of 2D Bi film in scanning tunneling microscopy experiments, providing a direct spectroscopic evidence of the TI nature. Furthermore, by combining angle-resolved photoemission spectroscopy and first-principles calculations, Miao *et al.*[24] realizes the epitaxial growth of 2D Bi(111) on $Bi_2Te_3$ substrate, but the topologically nontrivial properties of Bi film are strongly influenced by the substrate, the band dispersion near $\Gamma$ point is drastically modified due to strong hybridizations at $Bi/Bi_2Te_3$ interface. Recently, Zhou *et al.*[25] predict a new QSH insulator with a bulk gap of 0.18 eV in a single-layer $Bi_4Br_4$, which could be exfoliated from its 3D bulk material due to the weakly-bonded layered structure. Its topological nature, however, has not been observed experimentally. For 2D carbon-based monolayers, besides the hexagonal honeycomb lattice, graphene,[4,5] the porous allotropes of carbon phase, e.g., graphenylene,[26] graphyne,[27] and T-lattice[28], have been proposed, demonstrating abundant electronic properties depending on their crystal structures. By replacing carbon atoms in these structures with transition-metal dichalcogenides, however, some of them can turn into 2D TIs, but with relatively small band gaps.[29-31] A nature question then arises: does QSH effect exist in this porous layered structure containing Bi atom and possessing high feasibility in experiment? Addressing this question will not only enrich the physics of 2D TIs but also pave new way for designing Bi-based topological materials for realistic applications.

In the present work, based on first-principles calculations, we study the geometry, energetic stability, and electronic properties of an atom-thick bismuth monolayer, which are called as bismuthylene, as shown in Fig. 1(a). The phonon spectrum and molecular dynamics (MD) simulations confirm that this structure is dynamic stable at room temperature, even up to 500K. The analyses of the electronic structures reveals that it is a native QSH insulator with a large-gap of 0.29 eV at $\Gamma$ point, larger than the buckled (0.2 eV) and flattened (0.2 eV) bismuth, $Bi_4Br_4$ (0.18 eV), and stanene (0.1 eV) films, also more stable energetically than these structures. The bismuthylene has a finite gap and robust band topology under strains within -6 – 5 % and electric fields up to 0.8 eV/Å. Furthermore, a tight-binding (TB) model is constructed to explain the low-energy physics behind band topology induced by SOC, demonstrating by gapless helical edge states. Finally, we propose a quantum well (QW) structure by sandwiching bismuthylene between two BN sheets and reveal that bismuthylene remains topologically nontrivial with a sizeable gap. Our results enrich the 2D family of QSH insulator and enable its potential room-temperature applications in spinctronic devices.

First-principles calculations are performed by using density functional theory (DFT) [32] methods as implemented in the Vienna Ab initio simulation package (VASP) [33]. The projector-augmented-wave (PAW) potential [34, 35], Perdew-Burke-Ernzerhof (PBE) exchange-correlation functional [36], and the plane-wave basis with a kinetic energy cutoff of 500 eV are employed. The Brillouin zone is sampled by using a 9×9×1 Gamma-centered Monkhorst-Pack grid. The vacuum space is set to 20 Å to minimize artificial interactions between neighboring slabs. During the structural optimization of bismuthylene, all atomic positions and lattice parameters are fully relaxed, and the maximum force allowed on each atom is less than 0.02 eV/ Å. SOC is included by a second vibrational procedure on a fully self-consistent basis. The screened exchange hybrid density functional by Heyd-Scuseria-Ernzerhof (HSE06) [37] is adopted to further correct the electronic structure. The phonon calculations are carried out by using the density functional perturbation theory as implemented in the PHONOPY code [38].

Figure 1(a) displays the atomic structure of 2D bismuthylene monolayer with twelve Bi atoms in the unit cell, which contains two six-membered rings connected by a four-membered Bi unit, crystallizing in the hexagonal space group $P_{6m}$ (No. 175), with $a = b = 11.93$ Å. This new structure resembles the recently experimental reported planner graphenylene [26], but with a buckled height $d = 1.74$ Å induced by unsaturated Bi-$p_z$ orbitals, as illustrated by side view in Fig. 1(a). Interestingly, its buckled height is similar with the one ($d = 1.71$ Å) of buckled Bi (111) film.[17, 18] The dynamic stability of 2D bismuthylene has been checked by phonon spectrum calculations. As seen in Fig. 1(b), all branches have positive frequencies and no imaginary phonon modes. Further, the MD simulations is also performed, demonstrating that the bismuthylene is stable even up to room temperature ($T = 400$ K), as illustrated in Fig. 1(c).

Previous works [39] have reported the existence of honeycomb lattice bismuth and confirmed that the buckled 2D Bi (111) are more stable energetically than flattened one ($f$-Bi), because $sp^2$-hybridization is hard to achieve. Here similar results hold true for 2D bismuthylene. To demonstrate their relative stability, in Fig. 1(d) we present the formation energy differences of bismuthylene with respect to buckled Bi (111) and $f$-Bi films. Interestingly, the bismuthylene has lower energy than other two films under strains from -10 to 15 %, the energy difference is 199 meV ($\Delta E_1$) and 45 meV ($\Delta E_2$) for Bi (111) and $f$-Bi, respectively. More recently, the stanene has been well produced experimentally under Bi$_2$Se$_3$ substrate, its thermal stability can be enhanced up to 250 ℃ superior to GeH film. [40] As compared with each other, the formation energy of bismuthylene can reach much as large as 439 meV ($\Delta E_3$). It is reasonable to expect that the bismuthylene monolayer is rather stable in these 2D films, thus it is feasible to experimental realization.

The electronic band structures of 2D bismuthylene along high symmetry lines in the Brillouin zone (BZ) are shown in Fig. 2. In the absence of SOC, it exhibits a semiconducting character with a direct gap of 0.49 eV at $\Gamma$ point, though the global indirect gap is 0.42 eV with the valence band maximum (VBM) at M point and conduction band minimum (CBM) at $\Gamma$ point. An orbit-projected analysis for the composition of the electronic states reveals that, in the vicinity of the Fermi level at $\Gamma$ point, the top of valence band arises mainly from $p_{x,y}$ orbitals, while the bottom of conduction band consists of $p_z$ orbital. Such a band alignment has also been observed in tetragonal Bi bilayer structure.[41] With SOC inclusion, the band structures in Fig. 2(b) reveal a drastic change as compared with the case of without SOC. The most pronounced feature is that the top of valence band in the vicinity of Fermi level display a Mexican-hat shape near $\Gamma$ in BZ, suggestive of the band inversion with a large inverted gap. To better understand the band evolution process, we present a brief schematic of atomic orbitals to interpret the band structure at $\Gamma$ point in Fig. 2(c). Considering that the states near the Fermi level are mainly from Bi - $p$ orbitals, we neglect the effect of other orbitals, and summarize in three stages (I), (II), and (III). In stage (I) the chemical bonding occurs between Bi atomic orbitals, which results in Bi - $p$ orbital near the Fermi level splits into the bonding and anti-bonding states, *i.e.*, $p^+_{x,y,z}$ and $p^-_{x,y,z}$, where the superscripts + and – represent the parities of corresponding states, respectively. After switching on the crystal field at stage (II), these $p$ orbitals split into both $p_{x,y}$ and $p_z$ orbitals due to the threefold axis symmetry, with $p_z$ shifting upward with respect to degenerate $p_{x,y}$ orbitals. In stage (III) when the SOC effect is taken into account, the strong SOC drives the $p^-_z$ and $p^+_{x,y}$ states exchanged at the Fermi level, leading to an clear band inversion. Noticeably, this mechanism is different from $ZrTe_5$/$HfTe_5$[42], 1T-$MX_2$[43,44], and 1T'-$MoX_2$[45], where the SOC does not change band order, instead of only opening a sizable gap at the Fermi level. To overcome possible underestimation of the band gap, we employ additional HSE06[37] to confirm the existence of band inversion, see Fig. S1 in the supplementary information. Noticeably, the nontrivial gap of bismuthylene is enlarged to 0.28 eV by considering HSE, which are larger than ones of buckled Bi (111) (0.2 eV), [17-18] $f$-Bi (0.2 eV), ZrTe$_5$ (0.1 eV) [42], as well as stanene (0.1 eV) [9]. The large nontrivial band gap of bismuthylene without chemical adsorption, or field effects, is very beneficial for the future experimental preparation and makes it highly adaptable in various application environments.

Since this structure has an inversion center, the topological indices $v$ can be easily calculated by examining the parities of occupied band at four time reversal invariant momentum (TRIM), *i.e*, one $\Gamma$ and three $M$ points in BZ, based on the Fu-Kane's formula [46]. As listed Table I, we find that all the points (0.5, 0.0), (0.0, 0.5), and (0.5, 0.5) have + parity, while (0.0, 0.0) point has − parity, yielding a nontrivially topological invariant $Z_2 = 1$. Further, an alternative way to identify band topology is to introduce the evolution of Wannier Center of Charges (WCCs)[47] to calculate $Z_2$ invariant, in which the Wannier functions (WFs) related with lattice vector $R$ can be written as:

$$|R,n\rangle = \frac{1}{2\pi} \int_{-\pi}^{\pi} dk e^{-ik(R-x)} |u_{nk}\rangle$$

Here, a WCC $\bar{x}_n$ can be defined as the mean value of $\langle 0n|\hat{X}|0n\rangle$, where the $\hat{X}$ is the position operator and $|0n\rangle$ is the state corresponding to a WF in the cell with $R = 0$. Then we can obtained

$$\bar{x}_n = \frac{i}{2\pi} \int_{-\pi}^{\pi} dk \langle u_{nk}|\partial_k|u_{nk}\rangle$$

Assuming that $\sum_\alpha \bar{x}_\alpha^S = \frac{1}{2\pi}\int_{BZ} A^S$ with $S = I$ or $II$, where summation in $\alpha$ represents the occupied states and $A$ is the Berry connection. So we have the format of $Z_2$ invariant:

$$Z_2 = \sum_\alpha [\bar{x}_\alpha^I(T/2) - \bar{x}_\alpha^{II}(T/2)] - \sum_\alpha [\bar{x}_\alpha^I(0) - \bar{x}_\alpha^{II}(0)]$$

The $Z_2$ invariant can be obtained by counting the even or odd number of crossings of any arbitrary horizontal reference line. As expected in Fig. 3(a), the evolution lines of Wannier centers cross the arbitrary reference line an odd number of times in the $k_z = 0$ plane. Furthermore, the topological invariant $Z_2$ of bismuthylene at higher temperatures (such as 100K, 200K, 300K, 400K) is also examined, similar results with $Z_2 = 1$ are obtained. As a consequence, the bismuthylene monolayer is indeed a 2D TI even working at high temperature ($T > 300$ K).

The low energy physics behind band topology of 2D bismuthylene can be further explored via a Slater-Koster TB model.[48] The effective Hamiltonian in the reciprocal apace can be expressed as follows:

$$H = \sum_{i,\alpha} \varepsilon_i^\alpha c_i^{\alpha+} c_i^\alpha + \sum_{<i,j>,\alpha,\beta} t_{i,j}^{\alpha,\beta} c_i^{\alpha+} c_j^\beta + \lambda <\vec{L} \bullet \vec{\sigma}>_{\alpha\beta}$$

where $\varepsilon_i^\alpha$, $c_i^{\alpha+}$ and $c_i^\alpha$ represent the on-site energy, creation, and annihilation operators of an electron at $\alpha$-orbit of the $i$-th atom. The $t_{i,j}^{\alpha,\beta}$ is the nearest-neighbor (NN) hopping parameter of an electron between an $\alpha$-orbital of $i$-th atom and $\beta$-orbital of j-th atom. $\vec{L}$ is the angular momentum operator and $\vec{\sigma}$ is the Pauli matrix, in which the matrix element is determined by the basis of atomic orbits ($p_x$, $p_y$, $p_z$) of the $i$-th atom. $\lambda$ is the 1NN SOC strength of the Bi atom. For 1NN and 2NN hopping terms, the hoping parameters are expressed as:

$$t_{i,j}^{p_x p_x} = V_{pp\sigma} \times \cos^2\theta + V_{pp\pi} \times \sin^2\theta$$
$$t_{i,j}^{p_x p_y} = (V_{pp\sigma} - V_{pp\pi}) \times \cos\theta \times \sin\varphi$$
$$t_{i,j}^{p_x p_z} = (V_{pp\sigma} - V_{pp\pi}) \times \cos\theta \times \cos\gamma$$
$$t_{i,j}^{p_y p_y} = V_{pp\sigma} \times \cos^2\varphi + V_{pp\pi} \times \sin^2\varphi$$
$$t_{i,j}^{p_y p_z} = (V_{pp\sigma} - V_{pp\pi}) \times \cos\varphi \times \cos\gamma$$
$$t_{i,j}^{p_z p_z} = V_{pp\sigma} \times \cos^2\gamma + V_{pp\pi} \times \sin^2\gamma$$

where $\theta$, $\varphi$ and $\gamma$ are the angles of the vector pointing from the $i$-th atom to the $j$-th atom with respect to $x$, $y$ and $z$ axis, respectively. By fitting the TB results with first-principles calculations, as shown in Fig. 4(a), we obtain the optimized parameters of $V_{pp\sigma} = 1.69$ eV and $V_{pp\pi} = -0.57$ eV for the σ and π orbits, while the 2NN hopping parameters are $V_{pp\sigma} = 0.78$ eV and $V_{pp\pi} = -0.28$ eV, respectively.

The most prominent feature is that the band structure is sensitive to the SOC

strength λ, as illustrated in Fig. 4(d). When increasing λ continually from 0, the band gap of monolayer bismuthnylene begins to suppresses, leading to a complete band-gap closure at a critical point (λ = 0.65), and then it is reopens as SOC further increases and approaches the full realistic value (λ = 1). The representative band structures at λ = 0.65 and λ = 0.8 are shown in Figs. 4(b) and 4(c), where the band-gap closure and reopening are illustrated. Obviously, the topologically nontrivial band order is indeed induced by on-site SOC strength, but the reversed band component at the Fermi level is originated from the hoping parameters between Bi-$p$ atoms.

The QSH effect in 2D bismuthylene should support an odd number of topologically protected gapless conducting edge states connecting the valence and conduction bands at certain K-points. Thus, we construct the maximally localized Wannier functions (MLWFs) [49] and fit a TB Hamiltonian with these functions for the zigzag-type nanoribbons. As the energy bands at the Fermi energy are predominantly determined by Bi-$p$ orbitals, the MLWFs are derived from atomic $p$-like orbitals and the TB parameters are determined from the MLWFs overlap matrix, and the calculated results are shown Fig. 3(b). One can explicitly see that each edge has a single pair of helical edge states in the bulk band gap and cross linearly at the $X$ point. These features further prove the nontrivial nature of this structure, in nice agreement with the analysis of the parity ($Z_2$ = 1). Remarkably, a sizeable bulk gap can stabilize the edge states against the interference of the thermally activated carriers, which is beneficial for observing room-temperature QSH effect.

Experimentally, the 2D films should be lie or grown on a substrate. It is reasonable to question how interfacial strain and charge transfer will affect the electronic and topological properties of bismuthylene. Here in Fig. 5(a) we show the variations of band gap as a function of strain $a$. One can see that the bulk band gap of bismuthylene are modified significantly *via* strain, but the nontrivial band topology is preserved under the strains of -6 – 5 % under considering HSE. Interestingly, a larger gap of 0.49 eV is obtained for -6 % suppression. One the other hand, the band structures of 2D bismuthylene in a perpendicular electric field (*E*-field) are checked *via* interfacial charge transfer effect, as illustrated in Fig. 5(b). The most significant effect of *E*-field is to split the band spin-degeneracy away from high symmetry momenta, due to the inversion-symmetry breaking. Notably, the nontrivial topology is preserved with $E$-field up to 0.8 eV/Å. Such an external $E$-field induced tunable band gap and robustness against interfacial charge transfer offer a promising route to engineer electronic properties for the benefit of spintronics.

Finally, we propose a bismuthylene/$h$-BN QW, since the BN sheet is an ideal substrate to assemble 2D stacked nanodevices in experiments.[50,51] Figure 6(a) displays geometrical structure of bismuthylene/(5×5)BN QW with bismuthylene sandwiched between two BN sheets, where the lattice mismatch between them is only 2.75 %. After full relaxation with van der Waals (vdW) forces [52], 2D bismuthylene remains original free-standing structure with a distance between adjacent layers of 3.33 Å, with a binding energy of -61 meV, suggesting a typical VdW heterostructure. Figure 6(c) shows the band structure with SOC. We find that the contributions of $h$-BN substrate locate far away from the Fermi level, the states around the Fermi level being dominantly determined by Bi-$p_{x,y}$ and Bi-$p_z$ states with an inverted band order, which is similar to the case of a free-standing monolayer. Accordingly, the QW remains nontrivial, which is further confirmed by the topological invariant $Z_2 = 1$, indicating a robustness of QSH effect against the substrate.

In summary, we propose a new 2D bismuthylene that is dynamically and thermally stable based on phonon spectrum and MD simulations. It is identified to be a native QSH insulator with a gap of 0.29 eV at $\Gamma$ point, which is larger than the buckled (0.2 eV) and flattened (0.2 eV) bismuth, $Bi_4Br_4$ (0.18 eV), as well as stanene (0.1 eV), also more stable energetically than these systems. The bismuthylene has a finite band gap and a nontrivial band topology for strains up to -6 − 5 % and $E$-field reaching to 0.8 eV/Å. A tight-binding model is constructed to explain the low-energy physics behind band topology induced by spin-orbit coupling. To probe the possibility of experimentally realizing this monolayer, we propose a bismuthylene/BN QW. Interestingly, the bismuthylene remains topologically nontrivial with a sizeable gap when sandwiched between BN sheets. These findings provide a new platform for realizing low-dissipation quantum electronics and spintronics devices.

───────────────────────

**Acknowledgments:** This work was supported by the National Natural Science Foundation of China (Grant No.11434006, 11274143, 61571210, and 11304121).

**Table I** Parities of occupied spin-degenerate bands at the TRIM points for Bismuthylene. Here, we display the parities of 30 occupied spin-degenerate bands for Bismuthylene. Positive and negative signs denote even and odd parities, respectively. The sign in parentheses is the product of the parity eigenvalues of the occupied spin-degenerate bands.

| $\Gamma_i$ | Parity of $\xi_{2n}$ of occupied bands | $\delta_i$ |
|---|---|---|
| (0.0, 0.0) | −+++−−+++−−+−+++−−−−−+−−−+++− | − |
| (0.5, 0.0) | +++++++−−−+++−−−−−+−−−+−−− | + |
| (0.0, 0.5) | +++++++−−−+++−−−−−+−−−+−−− | + |
| (0.5, 0.5) | ++−−++−−+++−−+++−−++−−++−−−−+ | + |
| Bismuthylene | **$Z_2$ topological invariant** | **ν = 1** |

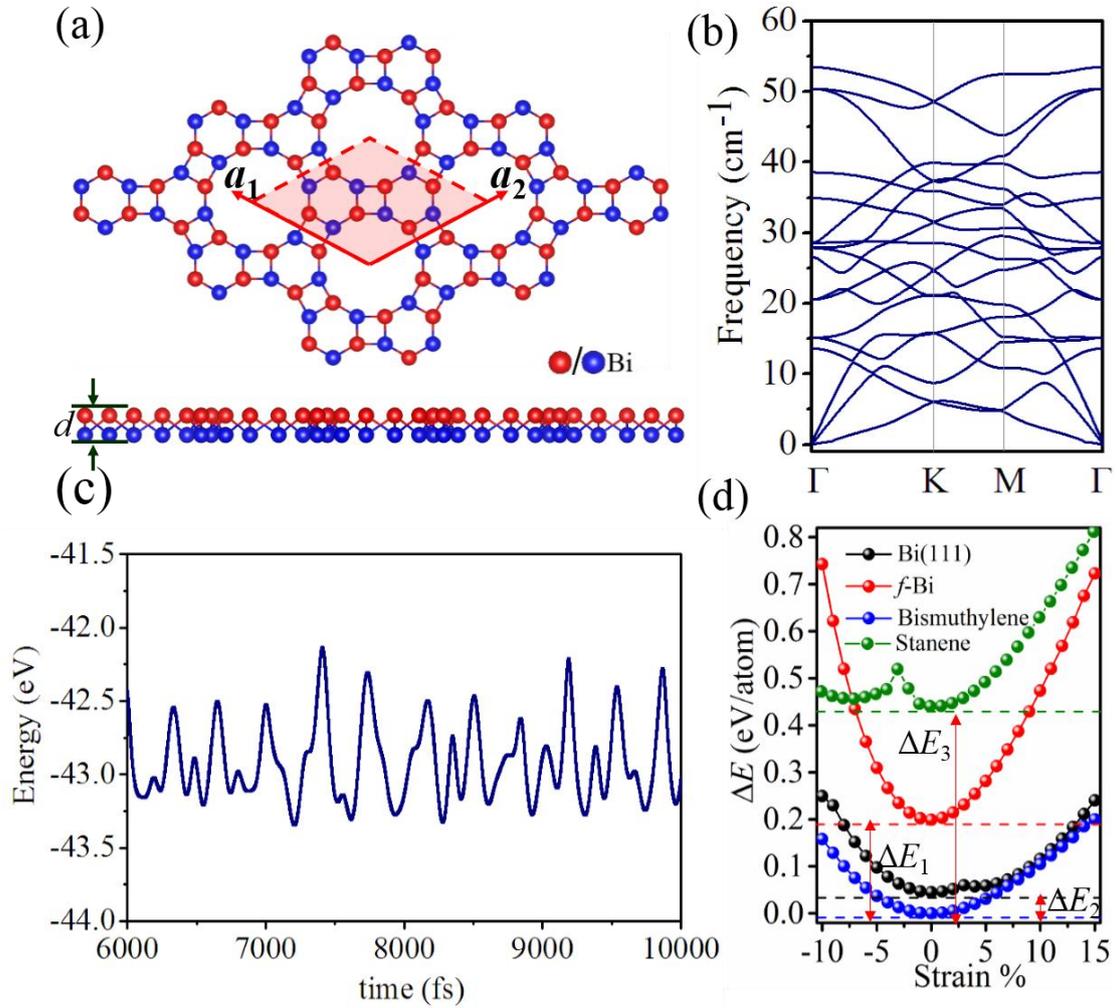

**Figure 1** (a) Top and side views of the geometrical structures of bismuthylene. Red and blue denote Bi upper and below atoms, respectively. Shadow area in (a) presents a unit cell. (b) Phonon band dispersion for bismuthylene. (c) Total energy fluctuations with respect to molecular dynamics simulation step for bismuthylene at 300 K with an Ab initio MD. (d) Energy of Bi(111), flattened Bi (*f*-Bi), bismuthylene and stanene under in-plane strain, $\Delta E_1$ = 199 meV/atom, $\Delta E_2$ = 45 meV/atom and $\Delta E_3$ = 439 meV/atom.

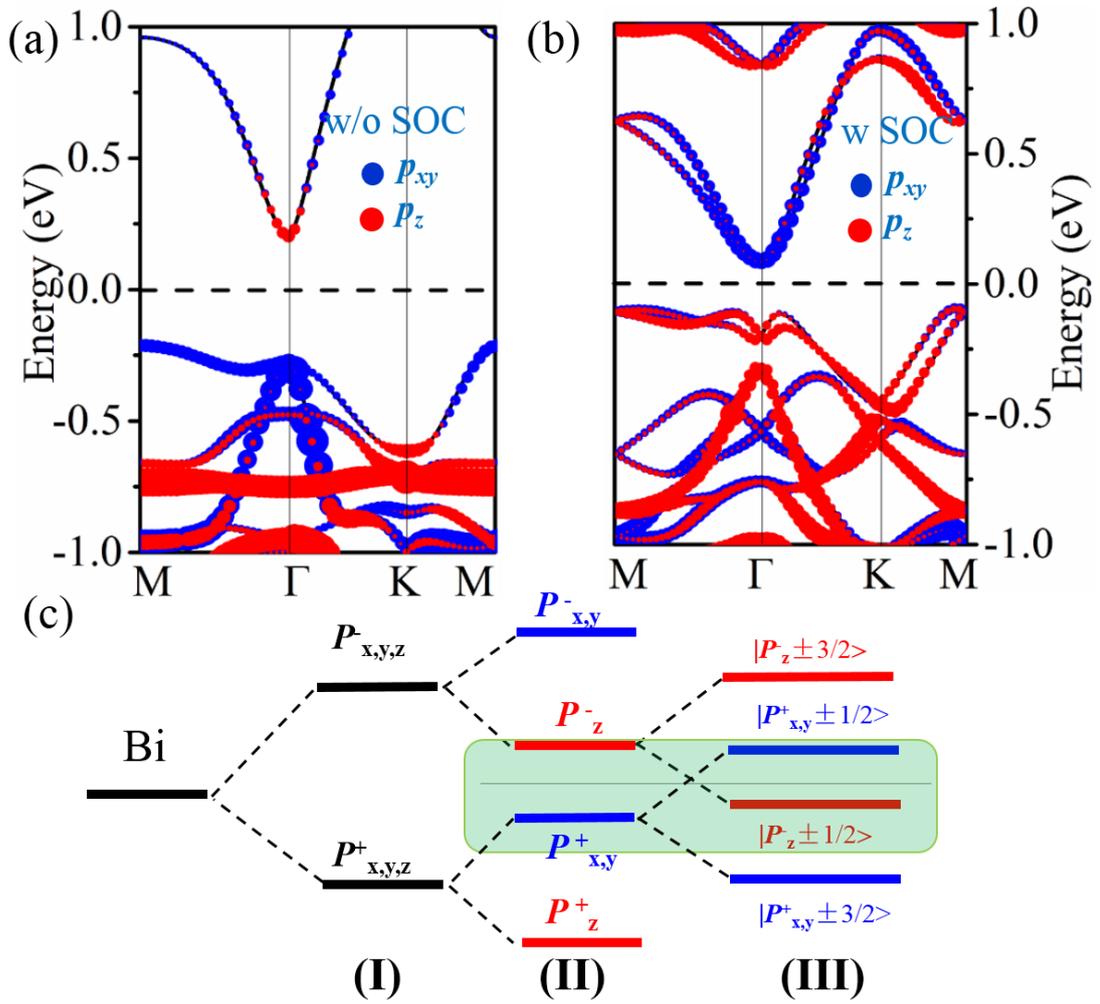

**Figure 2** The calculated orbital-resolved band structures and of bismuthylene (a) without and (b) with the spin-orbit coupling (SOC) effect, respectively. The blue and red dots represent the contributions from the $p_{x,y}$ and $p_z$ atomic orbitals of Bi atom. (c) The evolution of the atomic $p_{x,y}$ and $p_z$ orbitals of bismuthylene into the band edges at the $\Gamma$ point is described as crystal field splitting and SOC is switched on in sequence.

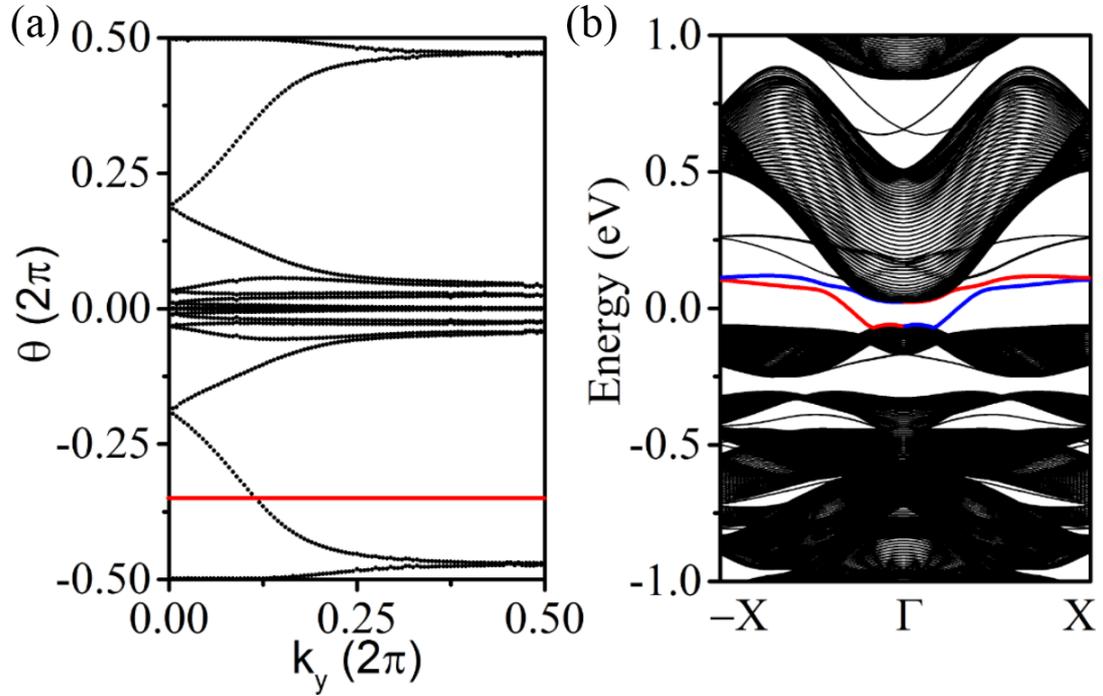

**Figure 3** Evolutions of Wannier centers along $k_y$ are presented in (a) bismuthylene. The evolution lines (blue dot lines) cross the arbitrary reference line (red line parallel to $k_y$) with an odd number of times, thus yielding $Z_2 = 1$. (b) Spin observed edge states of semi-infinite zigzag bismuthylene ribbon. Black lines illustrate the bulk bands, and red/blue lines denote the spin up/down helical edge states.

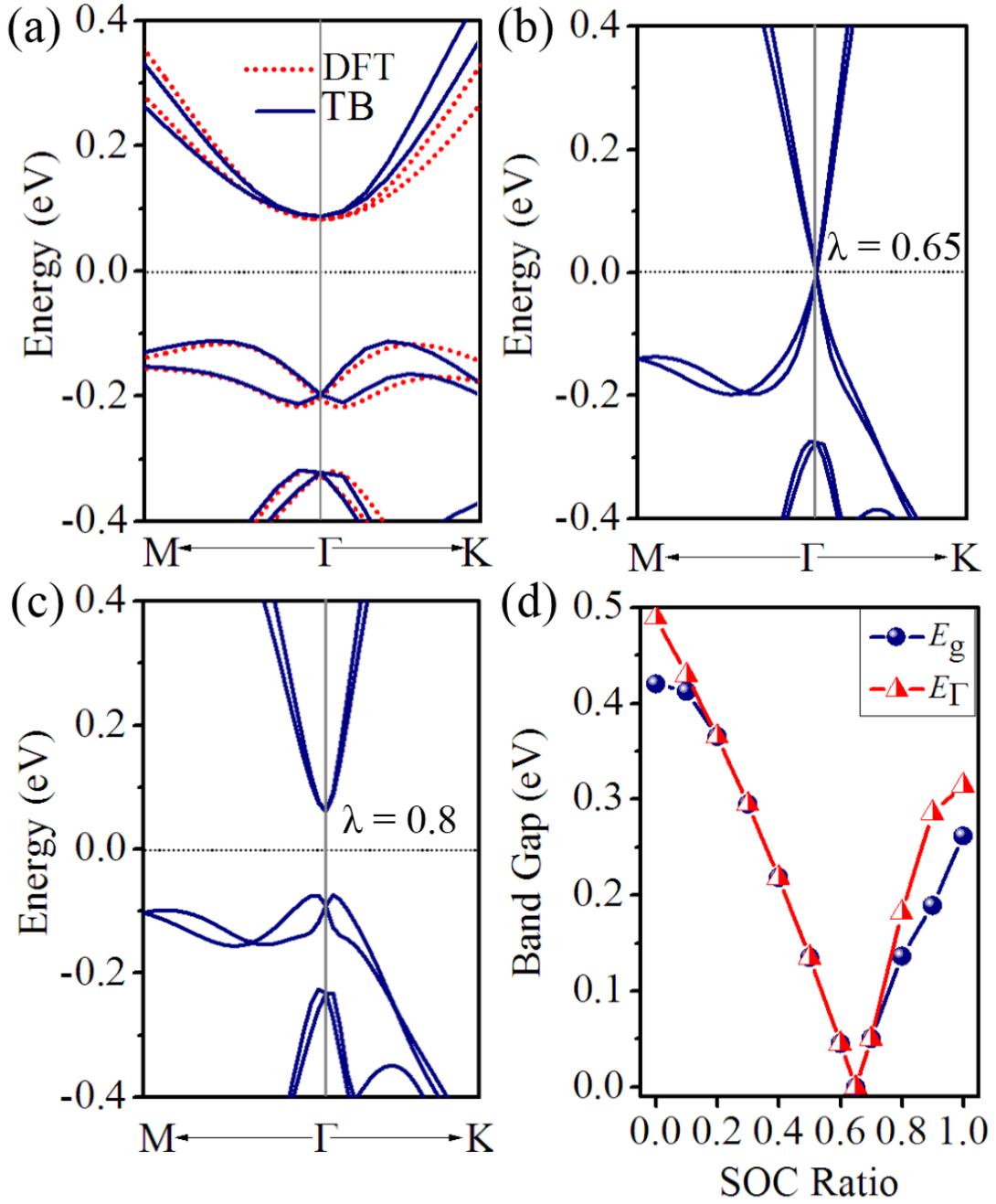

**Figure 4** (a) The electronic band structures from the tight-binding and first-principles calculations with the SOC. (b) and (c) the calculated orbital-resolved band structures of bismuthylene at three representative relative SOC coupling of 0.65 and 0.80, respectively. (d) The energy gaps at $\Gamma$ point and global energy gaps ($E$g) for bismuthylene upon varying the spin-orbit coupling strength from zero to the true value set as 1.

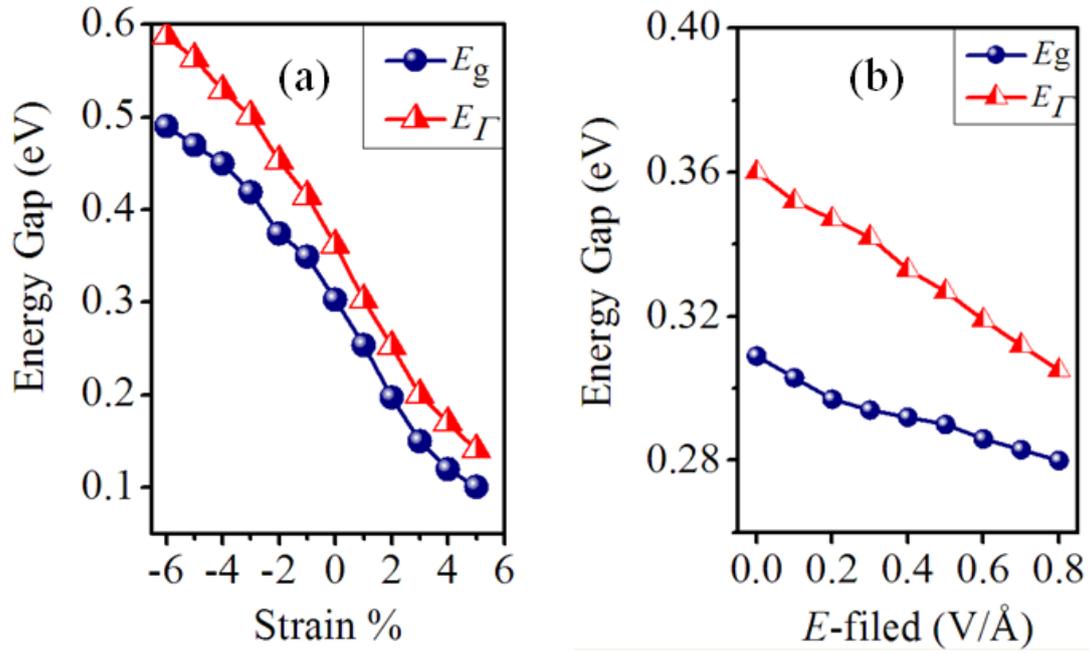

**Figure 5** (a) The calculated energy gaps at Γ point ($E_\Gamma$) and the global energy gap ($E_g$) of bismuthylene with SOC as a function of external strain. (b) and (c) show the calculated orbital-resolved band structures of bismuthylene under the tensile strain 3% and 6%, respectively.

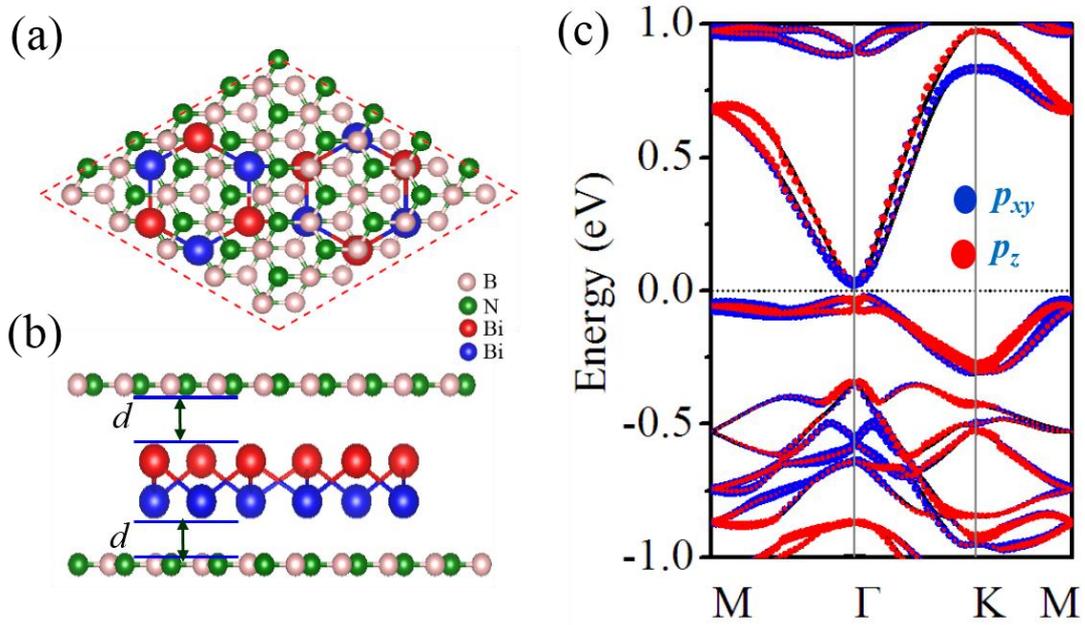

**Figure 6** (a) and (b) present crystal structures of bismuthylene grown on BN substrate from the top and side view. (c) and (d) bismuthylene@BN QW correspond to the band structure and orbital-resolved band structure with SOC. The blue and red dots represent the contributions from the $p_{x,y}$ and $p_z$ atomic orbitals of Bi atom.